\documentclass[10pt, conference, compsocconf]{IEEEtran}
\usepackage{color}
\usepackage[utf8]{inputenc}
\usepackage[T1]{fontenc}
\usepackage{cite}
\usepackage{fixltx2e}
\usepackage{graphicx}
\usepackage{multirow}
\usepackage{url}

\begin{document}

\title{The Metadata Anonymization Toolkit}
\author{\IEEEauthorblockN{Julien Voisin}
    \IEEEauthorblockA{Belfort-Montbéliard University Institute\\ of Technology (UTMB), France\\
    julien.voisin@dustri.org\\}
\and
    \IEEEauthorblockN{Christophe Guyeux, Jacques M. Bahi}
    \IEEEauthorblockA{Computer Science Laboratory DISC\\
FEMTO-ST Institute, UMR 6174 CNRS\\
University of Franche-Comt\'{e}\\
Besan\c con, France\\
\{jacques.bahi, christophe.guyeux\}@femto-st.fr}
}

\maketitle

\begin{abstract}
This document summarizes the experience of Julien Voisin during the
2011 edition of the well-known \emph{Google Summer of Code}.
This project is a first step in the domain of metadata anonymization in Free Software.
This article is articulated in three parts.
First, a state of the art and a categorization of usual metadata,
then the privacy policy is exposed/discussed in order to find
the right balance between information lost and privacy enhancement.
Finally, the specification of the Metadata Anonymization Toolkit (MAT) is presented,
and future possible works are sketched.

\end{abstract}

\section{Introduction}

The \emph{Google Summer of Code} (GSoC~\cite{GSoC}) is an annual program organized by Google,
to promote the development of free software:
students are paid by \emph{Google} to work on open-source projects during the summer break.
Julien Voisin, principal author of this paper, is an undergraduate student in computer
science at the University of Technology of Belfort-Montb\'{e}liard (University
of Franche-Comt\'{e}) who was selected
in the 2011 GSoC session. This article, written under the supervision of
the other authors, summarizes his experiences.

The project was to depersonalize or remove any personal information
embedded in media sent through the \emph{Tor} network~\cite{Tor}.
\emph{Tor}, which stands for \emph{The Onion Router},
 is a network of virtual tunnels that allows people and groups
to improve their privacy and security on the Internet.
It also enables software developers to create new communication
tools with built-in privacy features.
\emph{Tor} provides the foundation for a range of applications
that allow organizations and individuals to share information
over public networks without compromising their privacy.
However, the desired anonymity can be broken by the data sent
through it, if it contains information
on the sender. For instance, metadata is frequently added
during several data acquisition processes, possibly without the
user's knowledge or consent. Obviously, such information can be
problematic when desiring to reach anonymity on the Internet.

Metadata consists of information that characterizes digital data
like Microsoft Word documents, pictures, music files, and so on.
In essence, metadata does provide an excellent source of information
on every facet of the data that can be characterized, like who
produced the document, when and where it was produced,
for which reasons, what the content of the media is, and so on.
Indeed, metadata within a file can reveal a lot of things about
the author of the document.
For instance, cameras record the date on which any picture
has been taken and which camera has been used.
Office documents like PDF or those produced by LibreOffice/Microsoft Office
automatically add authors and company information
into documents and spreadsheets.
Not everybody is willing to disclose such information on the web.

This paper presents the main achievement of a project realized by Julien Voisin during
his 2011 session of the Google Summer of Code (GSoC).
This project aimed at removing any personal information embedded in any given media,
leading to the development of a Metadata Anonymization Toolkit (MAT)
library\footnote{The MAT is available at {https://mat.boum.org} and is distributed under the GPLv2 license.
Although it offers a command line interface, the most common usage is to use it by the mean of its GUI.}.
In this document, the MAT is presented, and the necessary choices that have been made during
its realization are documented. For instance, such a tool is supposes to be able to choose
which information to remove.
However, to determine which metadata raises problems for privacy is not a trivial question.
Some metadata is clearly problematic, such as GPS coordinates,
but others are less easy to tell apart (like the \emph{gid} (Group
IDentifier) giving the group of the file under consideration).
To solve this problem, the following politic has been chosen:
most of the time, the MAT library tries to remove all metadata
that is not mandatory to the file integrity.

The Metadata Anonymization Toolkit is already embedded in the Tails GNU/Linux distribution~\cite{tails}.
Tails is a live CD or live USB that aims at preserving the users privacy and anonymity in a friendly way.
More specifically, Tails delivers the three following services. Firstly, it
helps to use the Internet anonymously, from any location and on any computer:
all connections to the Internet are forced to go through the Tor network.
Secondly, it leaves no trace on the computer unless the user asks it explicitly.
Lastly, up-to-date cryptographic tools can be used to encrypt files, emails, and instant messaging.
Tails is not the only LiveCD dedicated to privacy.
Incognito~\cite{incognito} was a similar project in all aspects,
but based on Gentoo instead of Debian. The main developer has abandoned
the project to join forces with the Tails team in 2010.
JonDo LiveCD~\cite{jondo} also deserves a mention, although it is more focused
on JonDonym (their home-made anonymity network) rather than than on Tor and Amnesic features.

\begin{table*}[t!]
\centering
\begin{tabular}{|c|c|}
\hline
FIELD & EXAMPLE \\
\hline
File Name                       & example.jpg\\
File Modification Date/Time     & 2012:03:02 16:20:52+01:00\\
File Type                       & JPEG\\
MIME Type                       & image/jpeg\\
JFIF Version                    & 1.02\\
Orientation                     & Horizontal (normal)\\
X Resolution                    & 72\\
Y Resolution                    & 72\\
Resolution Unit                 & inches\\
Software                        & Adobe Photoshop CS3 Windows\\
Modify Date                     & 2009:06:14 21:09:57\\
Exif Image Width                & 1984\\
Exif Image Height               & 5300\\
Compression                     & JPEG (old-style)\\
Thumbnail Offset                & 332\\
Thumbnail Length                & 7172\\
Current IPTC Digest             & e8f15cf32fc118a1a27b67adc564d5ba\\
IPTC Digest                     & e8f15cf32fc118a1a27b67adc564d5ba\\
Copyright Flag                  & False\\
Photoshop Thumbnail             & (Binary data 7172 bytes, use -b option to extract)\\
Photoshop Quality               & 5\\
XMP Toolkit                     & Adobe XMP Core 4.1-c036 46.276720, Mon Feb 19 2007 22:40:08\\
Creator Tool                    & Adobe Photoshop CS3 Windows\\
Create Date                     & 2009:06:14 21:09:57+02:00\\
Metadata Date                   & 2009:06:14 21:09:57+02:00\\
Document ID                     & uuid:97B34EC31559DE1181FD86CC9CA57AAA\\
History                         & (empty)\\
Primary Platform                & Microsoft Corporation\\
Image Size                      & 1984x5300\\
Thumbnail Image                 & (Binary data 7172 bytes, use -b option to extract)\\
\hline
\end{tabular}
\caption{Example of metadata embedded in a picture (recovered with ExifTool~\cite{ExifTool})}
\label{table:metadata}
\end{table*}

The remainder of this student research work is organized as follows. In
Section~\ref{sec:metadataComposition}, the metadata under concern is
introduced. Section~\ref{sec:state-of-the-art} presents a short
state-of-the-art in current metadata forensics investigation.
Section~\ref{Theoretical Approach} analyses the different kinds of metadata
and proposes to categorize them.
Section~\ref{sec:MAT} presents the MAT tool, focusing on the chosen privacy policy
and implemented features, whereas in Section~\ref{sec:conclusion} future potential
improvements are addressed.
This research work ends by a conclusion section, where the contribution
in the field of anonymity is summarized and intended future work is
presented.

\section{Metadata Overview}
\label{sec:metadataComposition}

\subsection{Presentation}

Metadata, also known as \emph{data about data}, is information
that characterizes or gives details on digital media like music, images, movies, Office files, etc.
There are two kinds of metadata: structural and descriptive.

On the one hand, structural metadata provides information about the internal structure of the file.
Such information, which is required to extract the content from the binary representation,
does not change for a given type of file.
For instance, digital cameras insert into each picture they produce: the date, the camera model,
the post-processing software used, and even, for some high-end models, the GPS coordinates
of the place where the images have been taken!
Office documents like PDF or LibreOffice/Microsoft Office generally contains authors,
operating system, company information, and even the history of revisions into each document.

On the other hand, descriptive metadata provides unnecessary information about the file or
the data content.
Its reason to be is to enrich the media with secondary additional information like comments,
creation date, and so on. Obviously, it can also compromise
the anonymity and privacy of users in a network context, and previous works in the
literature show that such descriptive metadata
can allow the authors to be tracked back~\cite{Buchholz2004,Castiglione2007}.
For example, two zip files of the same size will have the same \emph{compressed size} metadata,
but not necessarily the same \emph{last modification date}.
More disquieting, some metadata is \emph{hidden}~\cite{Ohmura05Pat,Davis07Pat}:
it is added during the data acquisition stage of the file creation process,
possibly without the user's knowledge nor agreement.
Not everybody is willing to accept the automatic presence of such
information whose existence or content is not always known or precisely documented.
Examples of such metadata are presented in the next subsection,
whereas in Sect.~\ref{Case_Studies} some well-known case studies where metadata
has helped to find the author of a document are shown.

\subsection{Example of metadata contents}

Table~\ref{table:metadata} shows an example of the kind of metadata that can
be found in a given picture. This metadata has been obtained using \emph{Exiftool},
a Perl library/CLI application written by Phil Harvey for reading/writing
metadata, that supports many file types~\cite{ExifTool}.
It is possible to deduce at least the following facts from these data:

The last file modification of the picture is 2012:03:02, but the internal
metadata shows that the picture has been modified on 2009:06:14, so the picture
may not have been created as such. The software used is Adobe Photoshop CS3
on Microsoft Windows, this is why there is a Photoshop thumbnail embedded in
the medium. The IPTC digest\footnote{IPTC is a filestructure/type of metadata, like XMP} is both embedded and re-calculated,
allowing a simple integrity test to be performed.

The absence of a copyright flag may indicate
that the picture was not created by a
professional. The XMP toolkit
is the metadata normally used by Adobe, and the date corresponds to the release
of Photoshop CS3 (April 2007). The file's creation date and that in the metadata are the
same, which may indicate that the picture was created and not modified subsequently.
The document ID allows the tracking of the document. The absence of history
is a little bit unsusual, it cannot help to establish a precise chronology
of the image's creation timeline. The primary platform reinforces the hypothese
of a Windows platform. Since the thumbnail also contains its own metadata,
it is an additional vector that may compromises anonymity.

\subsection{Metadata against anonymity: some case studies}
\label{Case_Studies}

Metadata is generally not purposefully inserted in order to reveal any user's information
to unauthorized observers, it is embedded with the purpose of enriching the
media. Metadata provides
additional features and services, both to the user and the applications he or she uses.
Thus, removing all the metadata that may potentially lead to personal information leaks can appear
as excessive: to refuse concrete services that
metadata offers for largely hypothetical threats may look unreasonable.
However, metadata has already been successfully used to discover
the hidden authors of unauthenticated digital documents
published on the Internet, as illustrated by the two following examples.

\begin{figure}
\centering
\includegraphics[scale=0.45]{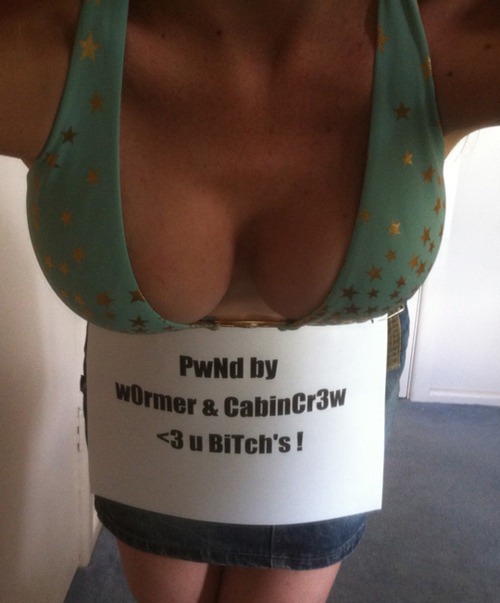}
\caption{Higinio's girlfriend (with GPS coordinates)}
\label{fig:groslolo}
\end{figure}

In February 2012, a hacker named Higinio (w0rmer) O. Ochoa III posted a link, on
his Twitter account, to a website disclosing data taken from various law enforcement websites.
On the bottom of the website was a picture of a woman's breast with a message
destined to mock the authorities (Fig.~\ref{fig:groslolo}).
Unfortunately for him, the photo was taken from an iPhone, which embeds GPS coordinates
among other metadata. These coordinates led the police to identify the girl in the picture,
who was the girlfriend of the hacker. He was sentenced to 27 months of prison~\cite{Dailymail2129257,CSO705170}.

Another well-known example of arrest using metadata is the case
of the infamous ``BTK serial killer'', namely Dennis Rader~\cite{abajournal}.
Some weeks before his arrest, he asked the police if he could communicate
with them using a floppy disk, without being traced back to a particular computer.
The police answered by posting an advertising in the local newspaper, as instructed by Rader, saying
``Rex, it will be OK''.
Some weeks later, such a disk was received at a local television station.
A forensic analysis (with the EnCase forensic software~\cite{EnCase}) revealed an erased Word document
that contained the terms ``Christ Lutheran Church'', and that was
last modified by ``Dennis''. Dennis Rader was the president
of the local church congregation’s council.
 The BTK had taken efforts to delete identifying information from the disk, but
 had printed the file in his church, because his printer was down.
The police lieutenant Ken Landwehr,  head of the multiagency task force in charge of the case,
said that ``this clue was a determinant factor for his arrest.
If he had just quit and kept his mouth shut, we might never have connected the dots.''

A more recent example is the capture of John McAfee, which was probably made possible
with the help of metadata forensics. Sought by the justice of Belize for
questioning about the murder of his neighbour, McAfee fled to Guatemala.
He was accompanied by two journalists from Vice
magazine, who took a picture of him during his escape.
Unfortunately for him, they forgot to wipe the metadata :
\begin{tabular}{lcl}
GPS Altitude & :& 7.1 m Above Sea Level\\
GPS Latitude                    &:& 15 deg 39' 29.40" N\\
GPS Longitude                   &:& 88 deg 59' 31.80" W\\
GPS Position                    &:& 15 deg 39' 29.40" N,\\
&&  88 deg 59' 31.80" W.
\end{tabular}

Such information probably led to his arrest by the Guatemalian authorities
for entering the country illegally.

Even if these case studies have concerned anonymity
disclosure of criminals demanded by police officers,
everybody should have in mind that such investigations can
be conducted illegally by dictatorial governments, or illegally by
non-authorized parties. The individual right to privacy is inalienable
as recalled by the UN Declaration of Human Rights,
the International Convenant on Civil and Political Rights,
and many other international and regional treaties.

\section{Digital Forensics and Anonymity: a State-of-the-art}
\label{sec:state-of-the-art}

Items of digital media are more easily modifiable than traditional media.
Due to the revolutions of personal computers and the Internet,
digital media is now of widespread use and can be found everywhere.
Various media manipulation programs have been developed over the last decades,
and they have been progressively simplified and are now accessible to everyone.
These programs and afferent consequences have necessitated the emergence of digital forensics,
a new discipline that flows initially from the need to address the challenges
arising from digital media manipulation.
An important goal of this recent discipline is to (in)validate the authenticity or integrity of
media, and it usually tries to answer the two following
questions~\cite{Dugelay11}.
Firstly, ``Who were the media producers?'' Secondly, ``Has the
media been manipulated, is it faked?''

Most of the multimedia forensics tools that try to addess these
questions work on the data themselves, as
removing the bullet-scratch of manipulation or creation is most of the time
a difficult signal processing task only accessible to experts~\cite{VanLanh07}.
In usual forensics frameworks, statistical signal processing analysis is
combined with information available through metadata, to respond
to these kind of questions~\cite{Herbert02}.
The best known software in forensic investigation,
which is nowadays a common law enforcement area, is maybe
EnCase~\cite{EnCase}, already referred to in this document.
Another well known tool is called COFEE (Computer Online Forensic Evidence Extractor~\cite{cofee}),
a software program developed by Microsoft to conduct forensic investigations,
which became famous when it was leaked on the Internet in November 2009. Microsoft
provides devices and free technical support to law enforcement agencies
all around the world. Although the majority of forensic software is closed-source and expensive,
some others are free, like the DEFT\footnote{http://deftlinux.net - DEFT is a
GNU/Linux liveCD running Windows forensic freeware using Wine}
 or CAINE\footnote{http://caine-live.net - A GNU/Linux forensic liveCD},
SLEUTH\footnote{http://sleuthkit.org - A Free library and collection of tools to
investigate disk images.}, and so on.

Oddly, the converse problem, namely metadata anonymization, is not a
well studied area of research in digital forensics.
The little that
has been investigated is about \emph{volunteer} anonymization, which happens
at the \emph{attacker} side.
For instance, Symantec has explored metadata anonymization for file system images~\cite{Symantech-report},
following an approach described below.
Symantec has its own file system called \emph{VxFS}. An interesting feature of this file system is its ability
to take a snapshot of metadata (no user data is retained), for forensic analysis
purposes, debugging, or troubleshooting.
If privacy is an important concern of the client, they can demand metadata anonymization:
all folder and file names will be hashed with the SHA-1 function.
Doing so is space and time efficient, and maintains the file system integrity
(the output of SHA-1 has a fixed length, so if the computed hash is too lengthy,
characters are removed, whereas it is simply processed by segment if this latter is too short).
Obviously, this rather simplistic approach can easily be bypassed,
but it prevents unintentional information leaks.
For Internet exchanges, the network utility \emph{Wireshark} allows an
anonymization of network packets,
whose headers can contain information about the transmitter and the receiver.
For more information, the reader is referred to~\cite{Wireshark-report} and the
references therein.

One can regret the relative absence of tools and research in the specific field of user's
metadata anonymization. This has been an important motivation to develop the MAT toolkit.
However, among the very few publications about metadata anonymization,
some significant research can be highlighted.
One of the most interesting papers concerns the
risks and countermeasures for PDF publication files~\cite{NSA-report}.
Although the approach focuses on \emph{Adobe Acrobat Professional},
this article explores a wide-range of potential attack vectors related to metadata,
and associated counter-measures.
The national library of New Zealand has also developed a tool called the ``Metadata Extraction Tool~\cite{Metadata}''
to obtain metadata from various document formats, in order to automatically
perform analysis or classification on them. It is entirely written in Java, and released under the
terms of the Apache license.

Anti-forensic measures have not yet been largely investigated.
Almost all public tools are too simplistic,
when not completely broken. But some are worth being mentioned like the MAFIA (Metasploit Anti-Forensic Investigation Arsenal),
which provides a suite of efficient tools, like Timestomp (NTFS timestamps removal).
Unfortunately, these tools are far too complex for the non-technical user.

\begin{table*}[t!]
\begin{tabular}{|c|c|c|l|l|c|}
\hline
\textbf{Type} & \textbf{Extension} & \textbf{Support} & \textbf{Metadata} & \textbf{Removal method} & \textbf{Remaining Metadata}\\
\hline
\hline
Portable&
&
&
&
Hachoir &
\\

Network&
.png &
Full &
Textual metadata + date &
or ExifTool  when available  &
None\\

 Graphics &
 &
&
 &
&
\\
\hline

Jpeg\footnote{Joint Photographic Experts Group} &
Jpeg  .jpeg, .jpg &
Partial &
Comment + exif/photoshop/adobe &
Hachoir or ExifTool if available &
Canon Raw tags\\
\hline

Open Document &
.odt, .odx, .ods, ... &
Full &
A meta.xml file contains all metadata &
Removal of the meta.xml file &
None\\
\hline

\multirow{2}{*}{Office Openxml} &
\multirow{2}{*}{.docx, .pptx, .xlsx, ...} &
\multirow{2}{*}{Full} &
A docProps folder containing  &
Removal of the docProps folder &
\multirow{2}{*}{None}\\

 &
 &
 &
xml metadata files &
 &
\\
\hline

 &
\multirow{6}{*}{.pdf} &
\multirow{6}{*}{Full} &
\multirow{6}{*}{A lot} &
Rendering of the PDF file on a &
None\\

 Portable &
 &
 &
 &
cairo surface with the help of &
\\

Document &
 &
 &
 &
 poppler, exporting the surface as &
\\

Fileformat &
 &
 &
 &
 a PDF. Metadata added during&
\\

 &
 &
 &
 &
 the process is removed with &
\\

 &
 &
 &
 &
 python-pdfrw. &
\\
\hline

\multirow{4}{*}{Tape ARchive} &
\multirow{4}{*}{.tar, .tar.bz2, .tar.gz} &
\multirow{4}{*}{Full} &
Tar metadata, metadata from &
Extracting and processing &
\multirow{4}{*}{None}\\

&
&
&
compressed files themselves, &
each file, recompression &
\\

&
&
&
metadata added by archive  &
in a new archive, processing  &
\\

&
&
&
to compressed files. &
of the new archive. &
\\
\hline

\multirow{4}{*}{Zip} &
\multirow{4}{*}{.zip} &
\multirow{4}{*}{Partial} &
Zip metadata, metadata from &
Extractiing and processing of  &
Metadata added \\

 &
 &
 &
compressed files themselves, &
each file, recompression in &
by zip itself \\

 &
 &
 &
metadata added by archive &
a new archive, processing &
into internal files\\

 &
 &
 &
to compressed files. &
of the new archive. &
\\\hline

\multirow{2}{*}{MPEG Audio} &
.mp3, .mp2, &
\multirow{2}{*}{Full} &
\multirow{2}{*}{id3} &
Hachoir &
\multirow{2}{*}{None}\\
&
.mp1, .mpa &
 &
 &
 or Mutagen when available &
\\
\hline

Ogg  &
.ogg &
Full &
Vorbis &
Hachoir  &
None\\

 Vorbis &
 &
 &
 &
or Mutagen when available &
\\
\hline

Free Lossless &
.flac &
Full &
Flac &
Mutagen &
None\\

 Audio Codec &
 &
 &
 Vorbis &
 &
\\
\hline

\end{tabular}
\caption{Supported file formats}
\label{table:les types}
\end{table*}

\section{Investigating Metadata}
\label{Theoretical Approach}
\subsection{Different types of metadata}

Items of metadata are of different kinds, depending on the way they are produced.
To have a clear understanding of their composition, we propose the following
classifications.

\subsubsection{The contextual type}

	Physical or logical devices that produce documents usually insert metadata into
	these produced files, mostly in order to enrich them.
	For instance, authors and software names are embedded in \emph{Office} documents,
	whereas author, interpreter, composer, and track number are usually inserted into multimedia data.
	They are not added to identify the user or producer, but still compromise their privacy.
	These types of metadata are generally documented, or can be easily obtained by reverse engineering.
	We will show that the MAT can handle them in most cases.

\subsubsection{The watermark type}

Some multimedia files embed a watermark directly in their data,
usually for copyright reasons.
If this data reveals no information about the user, it is
categorized into this ``watermark type''.
The authors are not aware of any case when
information susceptible to revealling the users identity was embedded
as a watermark

\subsubsection{The fingerprinting type}

For transmission of confidential documents, \emph{fingerprinting} metadata
(also known as a \emph{tattoos}) is often
embedded inside the document, to enable traceing back an eventual leaker.
Contrary to the watermark type, this metadata potentially
reveals some information about the file user.
Due to its final purpose, this type of metadata is really difficult to detect and
even more difficult to remove without breaking file integrity.
There are two sorts of fingerprinting metadata: robust, and fragile.

\paragraph{Robust fingerprinting}
The goal of robust fingerprinting is to insert hidden information
that resists content modification such as format conversion or resizing.
Robust fingerprinting is usually designed to be only removable
by modifying the content enough that it becomes completely unusable.

\paragraph{Fragile fingerprinting}
The fragile type of fingerprinting is usually used to guarantee the integrity of a document.
The slightest modification of the file will break the watermark.
The goal is to prove that the file has not been altered or modified.

Most of the proposed fingerprint removal solutions require
a large database of watermarked/non-watermarked files.
They imply complex statistical signal processing methods
coupled with artificial intelligence tools like
massive fine-tuned learning machines.\\
The removal of a fingerprint from a single document
remains an open problem, which is out of the scope of
this paper.

\section{Threat model}
The Metadata Anonymisation Toolkit adversary has a number of goals,
capabilities, and counter-attack types that can be used to guide us towards a set of
requirements for the MAT.
\subsection{Adversary}
\subsubsection{Goals}
\begin{itemize}
    \item Identifying the source of the document, since a
        document always has one. Who/where/when/how was a picture taken, where was
        the document leaked from and by whom, \dots{}
    \item Identify the author; in some cases documents may be anonymously authored
        or created. In these cases, identifying the author is the goal.
    \item Identify the equipment/software used. If the attacker fails to directly
        identify the author and/or source, his next goal is to determine the
        source of the equipment used to produce, copy, and transmit the document.
        This can include the model of camera used to take a photo, or which software
        was used to produce an office document.
\end{itemize}
\subsubsection{Positioning}
\begin{itemize}
    \item The adversary created the document specifically for this user.
        This is the strongest position for the adversary to have. In this case, the
        adversary is capable of inserting arbitrary, custom watermarks specifically
        for tracking the user. In general, MAT cannot defend against this adversary,
        but we list it for completeness.

    \item The adversary created the document for a group of users.
        In this case, the adversary knows that they attempted to limit
        distribution to a specific group of users. They may or may not have
        watermarked the document for these users, but they certainly know the
        format used.

    \item The adversary did not create the document, the weakest position for the adversary to have.
        The file format is (most of the time) standard, nothing custom is added: MAT should
        be able to remove all meta-information from the file.
\end{itemize}
\subsection{Requirements}
\subsubsection{Processing}
\begin{itemize}
    \item The MAT should avoid interactions with information. Its goal is to remove
        metadata, and the user is solely responsible for the information of the file.
    \item The MAT must warn when encountering an unknown format. For example, in
        a zipfile, if MAT encounters an unknown format, it should warn the user,
        and ask if the file should be added to the anonymized archive that is
        produced.
    \item The MAT must not add metadata, since its purpose is to anonymize files:
        every added items of metadata decreases anonymity.
    \item The MAT must handle unknown/hidden metadata fields, like proprietary
        extensions of open formats.
\end{itemize}

\section{Privacy and Anonymity Policy Implemented in the MAT}
\label{sec:MAT}

\subsection{Technical aspects}
MAT stands for Metadata Anonymization Toolkit. It is designed to
improve anonymity of files published online.
It consists of an extensible library, a Command Line Interface (CLI),
and a Graphic User Interface (GUI).
The MAT suite aims at providing, within the reach of anyone,
software dedicated to listing and removing metadata;
for portability purposes, it is entirely written in \emph{Python}~\cite{python},
and is based on the \emph{Hachoir} library~\cite{hachoir}.

\subsection{Security and anonymity policy}
To offer a reliable tool for metadata removal, one
must first determine criteria to decide whether a given field
must be considered harmful or not.
This raises the question of choosing a security and anonymity policy.
The strategy used by the \emph{MAT} is to process all the
metadata that can be removed: any piece of the file
that (1) is not a data, and (2) can be removed, is considered
as a threat and so is deleted.

This may seems rough, but it appears to the author of MAT as the best solution:
categorizing any possible metadata of every handled format
is, on the one hand, very subjective (for instance, is the
\emph{gid} field of a file a compromising metadata?),
and on the other hand it is an intractable task in practice.
Additionally, doing so leaves the least possible amount of metadata to the attacker.
Even if the absence of any metadata also provides information,
the quantity of information leaked by this absence is
obviously lower than the quantity provided by remaining metadata.

\subsection{White list approach}

Since the \emph{MAT} handles ``usual'' file formats, they are
often documented ones, even in the case of closed formats.
Thus, a white list approach is possible.
Because the format structure is known, each unknown
field is a non-standard one that can be safely
removed without breaking the file integrity.

A counterpart of this approach is that
some information loss may occur if the file
is not well-documented, or if it has been saved
using a non standard extension.
For example, Adobe use their own extension for PDFs.
So files produced with Adobe products, processed by the MAT,
and finally read with any Adobe products, may lose
information during this process.

Unfortunately, some closed formats are too complex to be completely
understood by a reverse engineering study, and so the MAT library
cannot handle them. This is the case for the Microsoft Office pre-2003 formats
($.doc$, $.ppt$, and so on),
which are known to be complex and whose design is often reported as disputable.

\subsection{Field Anonymization}

Metadata fields are suppressed whenever possible.
Otherwise, numerical data is set to 0, dates to Epoch,
and strings to an empty string.
Filling fields with random values or real-looking ones
may seem to make sense. But this is a poor strategy,
as producing false data is not harmless. It requires an
ad-hoc algorithm that could be traced back to
the owner or designer of this tailored anonymizing algorithm.
Additionally, potential input data of this algorithm
(seeds, PRNG parameters, and so on) can
accidentally leak information.

Indeed, except the cryptographically secure ones (like ISAAC or
BBS, see~\cite{Jenkins96,BBS, bfgw11:ij,bcgw11:ip}),
all the commonly used pseudorandom number generators (PRNGs) are quite biased.
Several attacks on common PRNGs have been reported~\cite{Arroyo08}, making it
possible to determine the algorithm used, and even
in some cases the seed, when considering a sufficiently large sample
of pseudorandom numbers generated by an algorithm.
So, using pseudorandom values instead of empty strings
may lead to a successful forensics attack, and to the discovery of some
tools used for this anonymization, revealing information on the user.
Furthermore, a simple source code inspection could allow an attacker to deduce
all possible values that the \emph{MAT} can generate,
allowing either brute force or probabilistic attacks.
Contrarily, removing everything that is
possible to remove is fairly safe.

\subsection{Surface Rendering for PDF}

The PDF format is quite complex: its specification document is
more than 1300 pages long. Furthermore major PDF producers have
developed their own extensions.
This is why such a format should be carefully cleaned part by part.

Of course, usual metadata is well documented. But
this format is so rich that it is possible to embed virtually anything
into a PDF: text, pictures, javascript, or even videos.
This is why the \emph{MAT} uses a clever trick to handle PDF: rendering on
cairo's PDF surface.
The rendering process is comparable to a print: hyperlinks are broken,
videos too, invisible metadata is removed, which drastically reduces
the risk of leaking the author's information.

\subsection{Supported formats and implemented features}

The \emph{MAT} supports most of the ``usual'' formats,
from pictures to Office documents (see Table~\ref{table:les types} for
an exhaustive list of supported formats).
Concerning archive formats (noting that most of the Office documents are
zipped XML files, and thus these formats belong into this category),
metadata mostly exists in a simple file folder,
which is easy to remove.
Audio format processing relies on Mutagen when available,
and images processing relies on Exiftool, again when available.

\subsection{The MAT output}

The MAT's output is intentionally minimal, since it is intended
for non-technical people. The goal of the "metadata listing" functionality
is to give a global view of present compromising metadata.
Since this is the same picture as the one studied previously,
we can recognize some patterns.

\section{Conclusion and Future Work}
\label{sec:conclusion}
This paper is a first step in the direction towards
document metadata processing for anonymity and privacy.
Based on a study of current trends in digital media forensics,
the problem of metadata removal has been pointed out as a potential vulnerability.
A thorough analysis of different metadata has allowed to propose a categorization
in two criteria: usefulness for media integrity,
and anonymity and privacy threats.
Finally, technical choices and their implementation have been presented.

Potential future improvements include dealing with fingerprints,
removing sensor ``bullet scratch'' from digital media, and processing more file formats.
The major short term improvement for the next version of the MAT is
the handling of metadata related to file-system-timestamps.
New file formats should be supported in the near future.
Furthermore, until now, the \emph{MAT} is neither able to detect nor to remove
fingerprints embedded in digital media.
This might not seem useful, as the insertion of
a watermark is usually a choice of user. However, more and
more camera models insert, in every captured image, a fingerprint
which identifies camera brand and model (see~\cite{Ohmura05Pat,Davis07Pat} and
therein references).


\bibliographystyle{IEEEtran}
\bibliography{mabase}

\end{document}